\documentclass{article}
\usepackage[latin2]{inputenc}
\usepackage[T1]{fontenc}
\usepackage{amsfonts}
\usepackage[dvips]{graphicx}
\usepackage{epic,eepic,pstcol}

\input{epsf}

\begin{document}
\title{\bf Chaotic self-similar wave maps coupled \\to gravity}
\author{Sebastian J. Szybka \\
{\small {\it Institute of Physics, Jagellonian University, Krak\'ow, Poland}}
\\
{\small {\it Institute for Theoretical Physics\footnote{preprint no. UWThPh-2003-29}, University of Vienna, Vienna, Austria}}
}
\maketitle
\begin{abstract}
We continue our studies of spherically symmetric self-similar
solutions in the $SU(2)$ sigma model coupled to gravity. For some
values of the coupling constant we present numerical evidence for the chaotic 
solution and the fractal threshold behavior. We explain this phenomenon 
in terms of horseshoe-like dynamics and heteroclinic intersections.
\end{abstract}
%\newpage
%\tableofcontents
%\newpage
\section{Introduction}
This is the fourth paper in a series aimed at understanding of the structure of
 the self-similar 
spherically symmetric wave maps coupled to gravity. In the first
two papers \cite{bizon2,bizon4} it was shown that for
the coupling constant $\alpha<\frac{1}{2}$ there exists a countable 
family of solutions that are analytic below the past light cone of 
the central singularity. It was also shown that there are two generic
types of solutions beyond the past light cone. The threshold between
these two generic types resembles many aspects of critical phenomenon
and has been studied in \cite{periodic_wm} for the coupling constant
$\alpha<0.426$. It was found that for this coupling the threshold
solutions are the periodic solutions.
Now, we extend studies of the threshold solutions for larger coupling constant
and examine the structure of the bifurcations in this parameter. This article
is mathematically self-contained but for physical background we refer the
reader to \cite{bizon2,vienna2}.
 
\section{Setting the problem}
Under the assumption of spherical symmetry and self-similarity the Einstein's
equations with wave map matter reduce to a system of 
autonomous \hbox{equations \cite{bizon2}}
\newpage 
\begin{eqnarray}
\label{eq:1} W'&=&-1+\alpha(1-W^{2})F'^{2},\\
\label{eq:2} A'&=&-2\alpha AWF'^{2},\\
\label{eq:3} (AF')'&=&\frac{\sin(2F)}{W^{2}-1},
\end{eqnarray}
subject to the constraint
\begin{equation}\label{eq:4}
1-A-2\alpha \sin^{2}(F)+\alpha AF'^{2}(W^{2}-1)=0.
\end{equation}
Physically the functions $W$ and $A$ parameterize the metric and the 
function $F$ parameterizes the corotational wave map. $\alpha$ stands for a 
dimensionless coupling constant ($\alpha=0$ corresponds to no gravity). 
We are interested in solutions of (\ref{eq:1}--\ref{eq:4}) starting 
at $x_{0}$ (we choose $x_{0}=0$) with the following initial conditions
\begin{equation}\label{eq:wp_final}
\begin{array}{llll}
W(0)=1,&A(0)=1-2\alpha,&F(0)=\frac{\pi}{2},&F'(0)=b
\end{array}
\end{equation}
where $b$ is a free parameter (since the system has reflection symmetry
$F\rightarrow -F$ we can take $b>0$ without loss of generality).
Solving (\ref{eq:4}) for $A$  
and using a new variable $D=F'$ we rewrite
equations (\ref{eq:1}--\ref{eq:3})
as the three dimensional, autonomous, first order system 
($\alpha<\frac{1}{2}$)
\begin{eqnarray}
\label{eq:1a} W'&=&-1+\alpha(1-W^{2})D^{2},\\
\label{eq:2a} D'&=&2\alpha WD^{3}+\frac{\sin(2F)}{-1+2\alpha {\sin(F)}^2}\left(\alpha D^2+\frac{1}{1-W^2}\right),\\
\label{eq:3a} F'&=&D.
\end{eqnarray}
with the initial conditions
\begin{equation}\label{eq:ic}
\begin{array}{lll}
W(0)=1,&D(0)=b,&F(0)=\frac{\pi}{2}.
\end{array}
\end{equation}
In the following
we will refer to solutions of (\ref{eq:1a}--\ref{eq:3a}) satisfying
the initial conditions (\ref{eq:ic}) as $b-orbits$.
It was shown in \cite{bizon4} that
for $\alpha\in[0,\frac{1}{2})$ $b-orbits$ exist locally and are analytic 
in $b$ and $x$.

\section{Types of solutions}\label{sec:st}
It was shown in \cite{bizon2} that for small values of $b$ the
$b-orbits$ exist up to the future light cone ($W=-1$) of central singularity,
while for large values of $b$ the $b-orbits$ develop a sort of
singular apparent horizon beyond which they cannot be continued ($W=+1$).
These two types of solutions, call them type A and type B
solutions, respectively, can be shown to form open sets \cite{periodic_wm}.
Hence, there must exists solutions which are not type A or B,
call them type C solutions. For a given $\alpha$ there exists
a critical value $b^{*}(\alpha)$ such that solutions with $b<b^{*}$
are of type A, solutions with $b>b^{*}$ are of type B, and the
solutions with  $b=b^{*}$
is of type C; in other words, we have a bistable behavior with
two generic final states A ($W=-1$) and B ($W=+1$), 
and the separatrix C ($|W|<1$ for all $x$).
We use the same terminology to
distinguish attractors A ($W=-1$) and B ($W=+1$). 

It was shown in \cite{periodic_wm} that type C solutions are
asymptotically periodic for $\alpha<0.426$. We will show in this
paper that this remains valid for $\alpha>0.482$.  However, we 
will also show that 
for $\alpha\in(0.426,0.482)$ (approximately) the transition between types
A and B does not occur at a single point but in a narrow interval
$(b^{*}_{min},b^{*}_{max})$ and the behavior of type C solutions is 
chaotic. This chaotic behavior is the main point of our interest.

\section{Type C solutions}

Numerical data show that for general initial conditions the whole phase 
space contains two basin sets (of attractors A and B) 
separated by a two dimensional surface. Type C solutions lay in this 
critical surface. 
In other words all initial points in the critical surface 
give type C solutions. The critical 
surface is an invariant submanifold.

\begin{figure}[!ht]
\begin{center}
\includegraphics[angle=0, width=12cm]{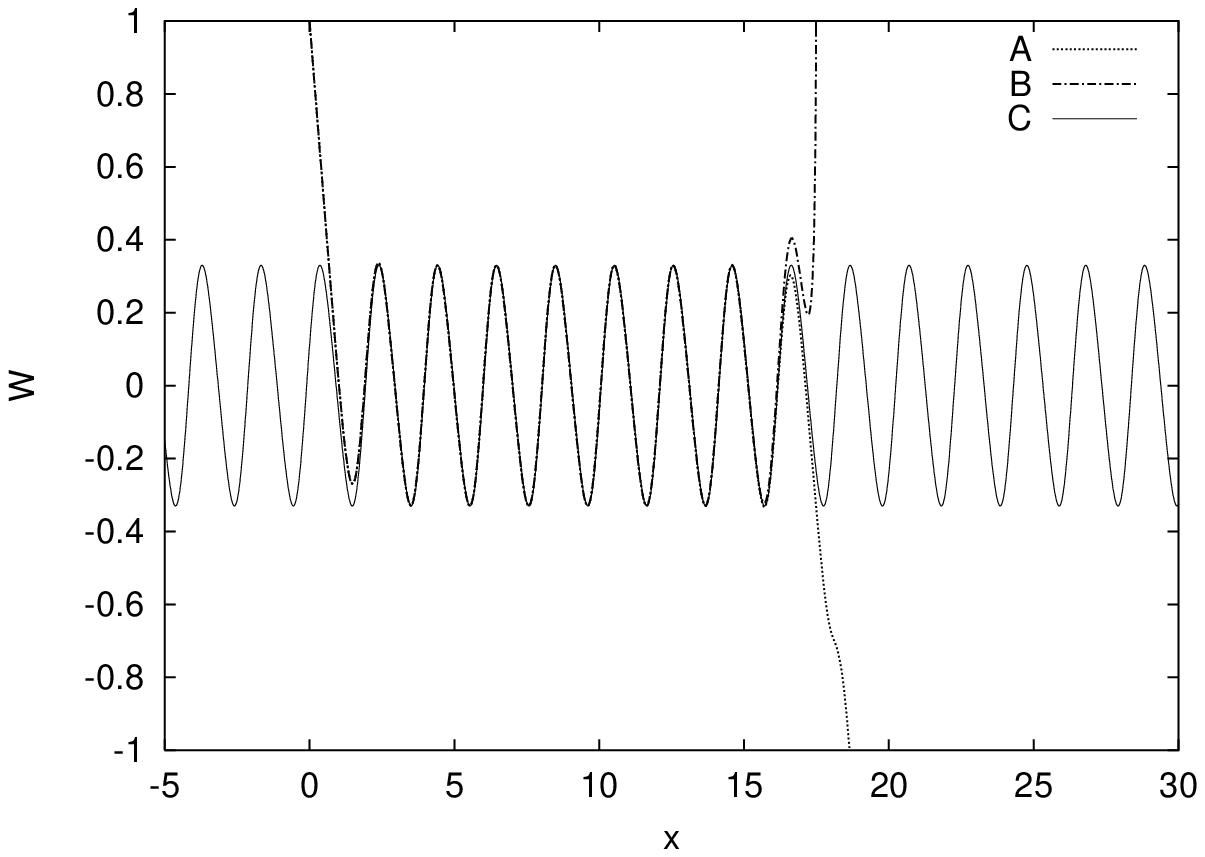}
\end{center}
\caption
{ The function $W$
- the type C solution (periodic).
The type A solution ($c=2.36134$), 
the type B solution ($c=2.36135$). 
All calculations for  $\alpha=0.38$. 
The type A, B solutions follow periodic C-attractor for
some time and then run towards A and B attractor.}
\label{fig:9a}
\end{figure}

We are especially interested in asymptotic structure of type C solutions.
Hereafter we will skip initial part of C solutions to simplify 
all figures. Let us call our asymptotic structure \hbox{C-attractor}.
In fact, it is the attractor of codimension one 
(two stable and one unstable directions).
The structure of \hbox{C-attractor} for $\alpha<0.426$ has been
found in \cite{periodic_wm} it is the periodic solution. In other words, 
$b-orbits$ starting 
close to the critical $b^{*}$ follow periodic C-attractor for some time and
then run towards A or B attractor (see figure \ref{fig:9a}).

\section{The bifurcation diagram}

\begin{figure}
\begin{center}
\includegraphics[angle=0, width=12cm]{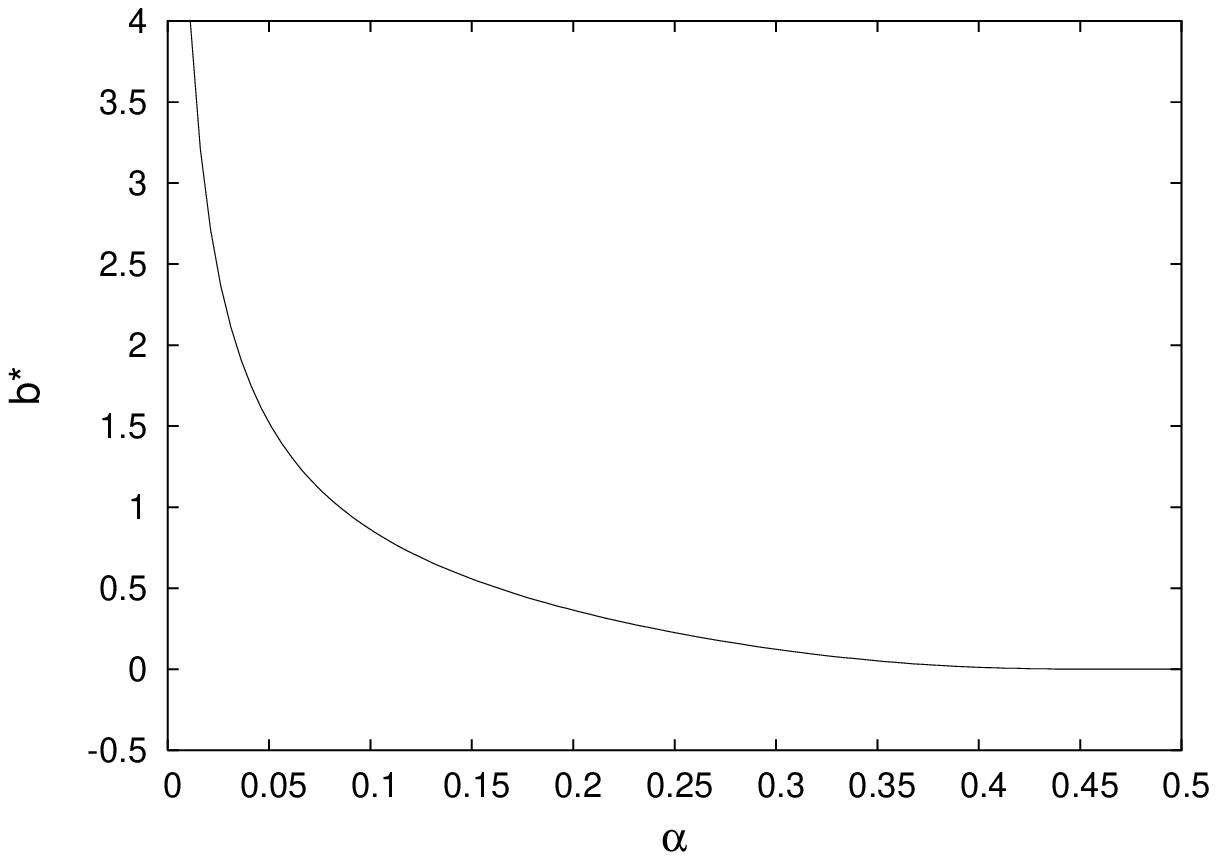}
\end{center}
\caption{The bifurcation diagram, lattice
$\Delta\alpha=10^{-3}$ i $\Delta b=2\cdot 10^{-3}$.
\hbox{The critical} curve $b^{*}(\alpha)$ separates
type A solutions from type B solutions.}
\label{fig:bif_b}
\end{figure}

\begin{figure}
\begin{center}
\includegraphics[angle=0, width=12cm]{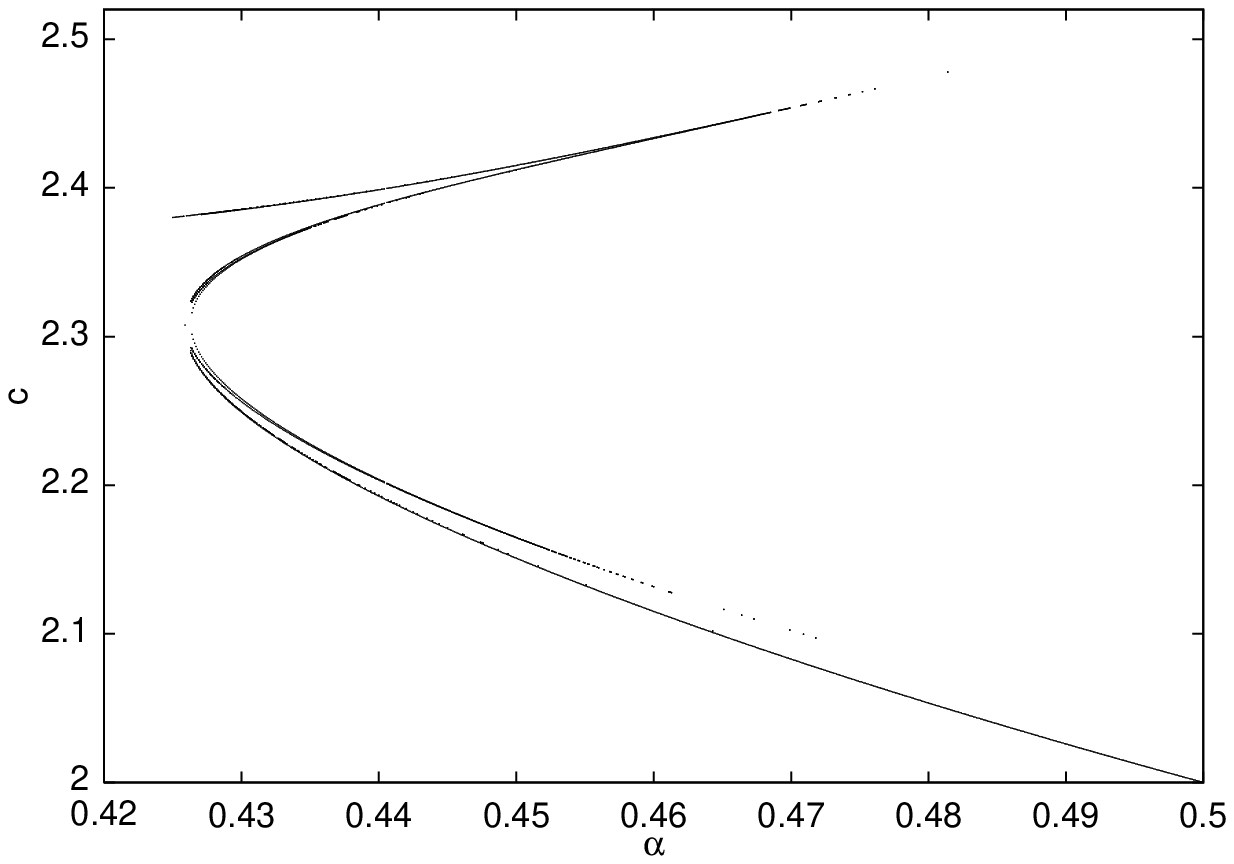}
\end{center}
\caption{The bifurcation diagram, lattice
$\Delta\alpha=10^{-4}$ i $\Delta c=2\cdot 10^{-4}$.
\hbox{The critical} curve $c(\alpha)$ separates
type A solutions from type B solutions.}
\label{fig:6}
\end{figure}

\begin{figure}
\begin{center}
\includegraphics[angle=0, width=12cm]{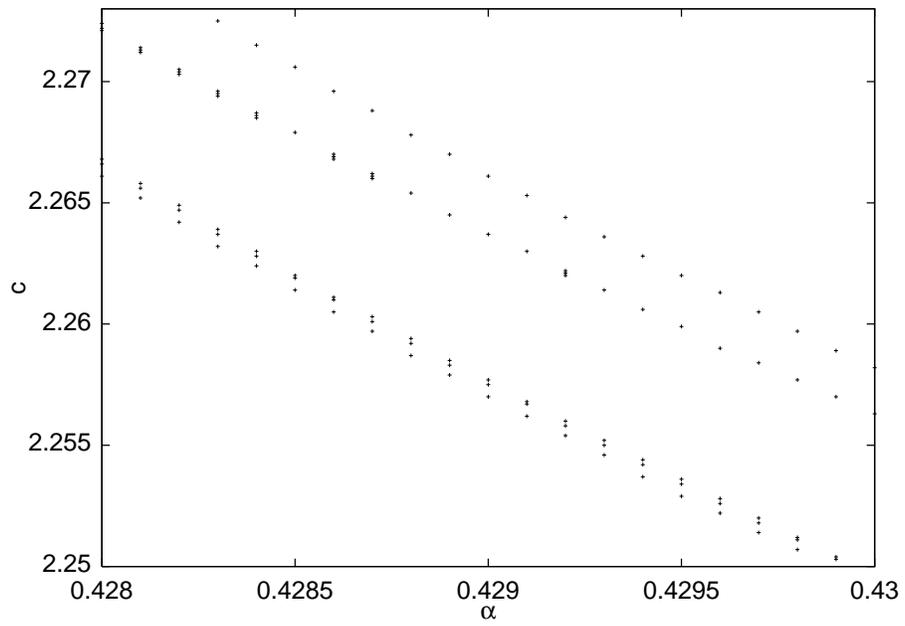}
\end{center}
\caption{The bifurcation diagram (figure \ref{fig:6})
enlarged 1000 times ,
lattice \hbox{$\Delta\alpha=10^{-4}$} and \hbox{$\Delta c=2\cdot
10^{-4}$}. The complex structure of the basin boundary 
is visible.}
 \label{fig:7a}
\end{figure}

\begin{figure}
\begin{center}
\includegraphics[angle=0, width=12cm]{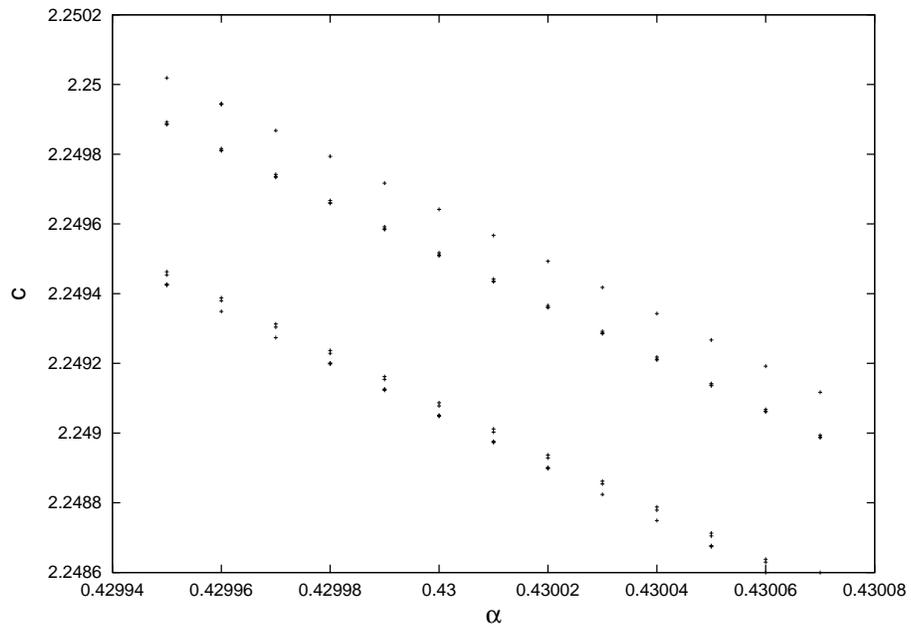}
\end{center}
\caption{The bifurcation diagram (figure \ref{fig:6}) 
enlarged 200000 times,
lattice $\Delta\alpha=10^{-5}$ i $\Delta b=10^{-6}$.
The complex structure of the basin boundary 
is visible.} \label{fig:7b}
\end{figure}

Let us consider the initial curve in the form (\ref{eq:ic}). 
For each $\alpha$, using the bisection 
procedure, we can find a critical values of $b^*$ parameter
which determine intersection of the initial curve with critical surface
(figure \ref{fig:bif_b}).
It has been noted in \cite{bizon2} that for large values of $\alpha\sim 0.43$,
$b^*(\alpha)$ is not a 
single point but there is a structure hidden in a narrow
interval $({b^*}_{min}(\alpha),{b^*}_{max}(\alpha))$. 
The solutions here are sensitive to
initial conditions infinitesimally small change of $b$ parameter can result
in a completely different final state. 
We have confirmed these numerical results.
Since the magnitude of interval 
$({b^*}_{min}(\alpha),{b^*}_{max}(\alpha))$
is of order $10^{-9}$ we used quadruple 
machine precision. In addition, the initial conditions (\ref{eq:ic}) are given
at a singular point.
We are interested only in asymptotic structure of type C solutions so
it is more convenient to reformulate
initial curve to start directly from \hbox{C-attractor}
\begin{equation}\label{eq:ic_new}
\begin{array}{lll}
W(0)=0,&D(0)=c,&F(0)=0,
\end{array}
\end{equation}
where $c$ is a function of $b$ such that 
$|{c^*}_{max}-{c^*}_{min}|>>|{b^*}_{max}-{b^*}_{min}|$.
The function $c(\alpha)$ is qualitatively similar to $b(\alpha)$ but now at
$\alpha$ approximately equal to $0.426$ we can resolve better the 
new structure 
in the diagram (figure \ref{fig:6}). 
This is a bifurcation point. Numerical results suggest
that this new structure is self-similar (figures \ref{fig:6}, \ref{fig:7a},
\ref{fig:7b}) so for 
each $\alpha$ we have 
infinitely many critical points. 
This suggests that the basin boundary is fractal.

There is not enough numerical data to calculate the fractal
dimension from the definition but
the phenomenon of fractal basin boundaries is well known 
so we have another tools to do this. We use the procedure suggested in
\cite{fractalbb} which goes as follows.

\begin{figure}
\begin{center}
\includegraphics[angle=-90, width=12cm]{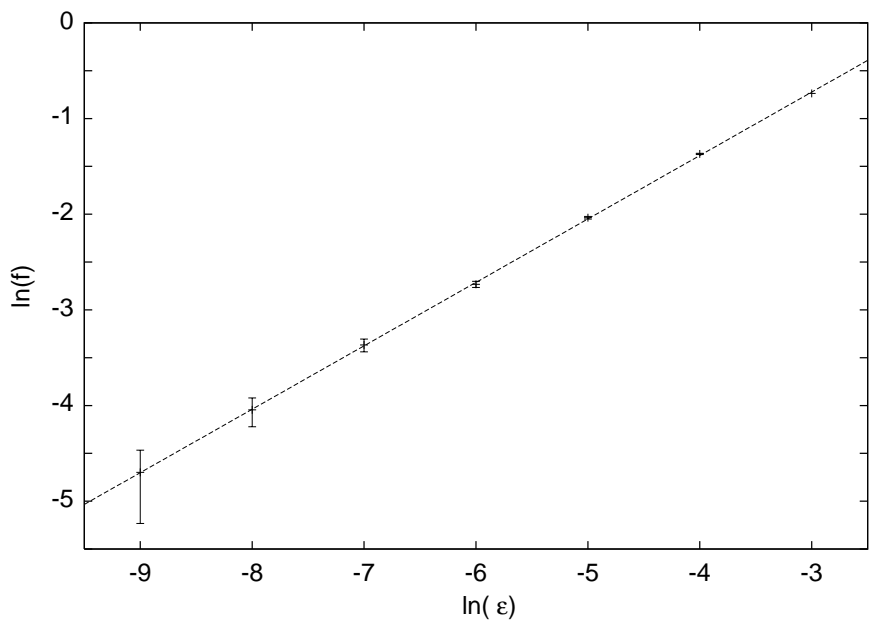}
\end{center}
\caption{The number of uncertain points in the phase space 
(for \hbox{$\alpha=0.4264$}) as
a function of the perturbation parameter $\epsilon$ (logarithmic
scale). The linear fit gives \hbox{the slope} \hbox{$a=0.663\pm 0.003$}
from which using (\ref{eq:fbb2}) and (\ref{eq:fbb0}) we get 
the capacity dimension \hbox{$d=0.337\pm0.003$}.} \label{fig:10}
\end{figure}

\begin{figure}
\begin{center}
\includegraphics[angle=0, width=12cm]{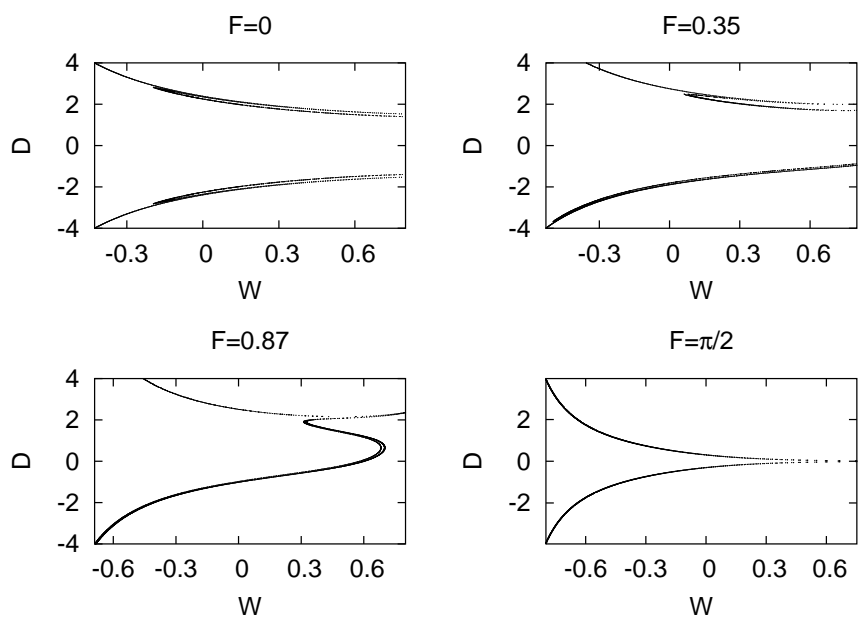}
\end{center}
\caption{The basin boundary for $\alpha=0.43$.
The complex structure is visible.}
\label{fig:4a}
\end{figure}

We call a point $\epsilon$-uncertain if perturbations with $\epsilon\ll 1$
in opposite directions
give points which belong to different basin sets.
It was shown in \cite{fractalbb} that the number of uncertain points in the 
phase space $f$ scales as
\begin{equation}
\label{eq:fbb1}f(\epsilon)\sim \epsilon^{a}.
\end{equation}
Therefore, in the limit we have
\begin{equation}
\label{eq:fbb2}
\lim_{\epsilon\rightarrow0}\frac{\ln f(\epsilon)}{\ln \epsilon}=a.
\end{equation}
It was shown in \cite{fractalbb} that the uncertainty exponent 
$a$ is related to the capacity 
dimension $d$ in $D$ dimensional phase space by the formula
\begin{equation}
{\label{eq:fbb0} d=D-a}.
\end{equation}
For $\alpha=0.4264$ 
we estimate numerically the capacity dimension to be \\ 
\hbox{$d=0.337\pm 0.003$}
(see figure \ref{fig:10}).
Since C-attractor lays in critical surface with infinite area
one should expect complicated dynamic.
Figure \ref{fig:6} shows only one dimensional relation. It would be
interesting to look at the whole basin boundary. We can do this by 
taking slices of the phase space (see figure \ref{fig:4a}).

\section{C-attractor}

\begin{figure}
\begin{center}
\includegraphics[angle=0, width=12cm]{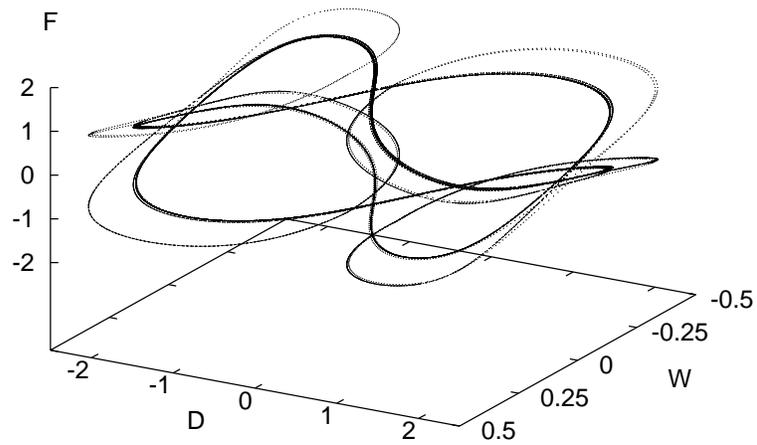}
\end{center}
\caption{\hbox{C-attractor}  in phase space
$(W,D,F)$ (for $F$ we identify the intervals 
$[-\frac{\pi}{2}+\pi k,\frac{\pi}{2}+\pi k]$ where
$k\in N$), \hbox{$\alpha=0.43$}.}
 \label{fig:1}
\end{figure}

\begin{figure}
\begin{center}
\includegraphics[angle=0, width=12cm]{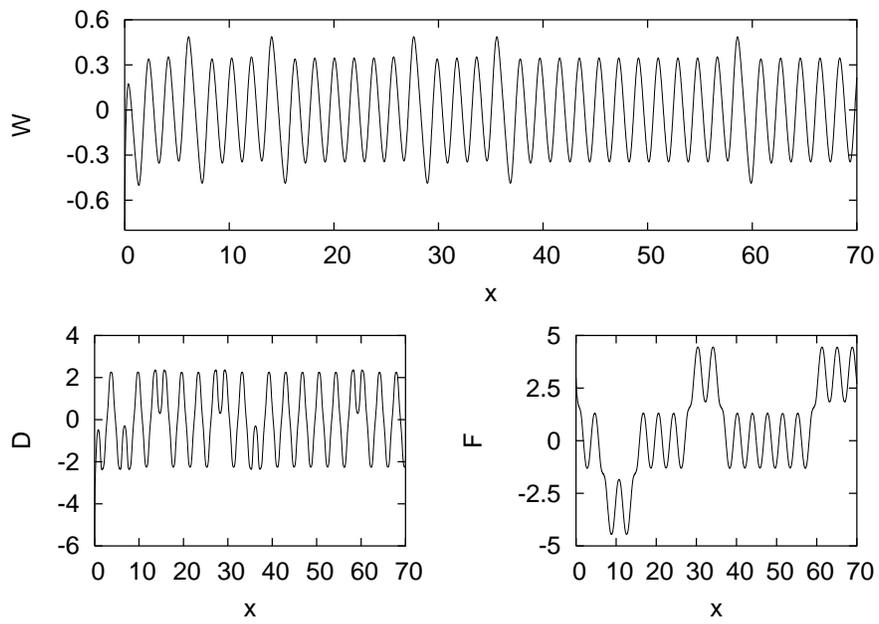}
\end{center}
\caption{The typical solution on \hbox{C-attractor}, \hbox{$\alpha=0.43$}}
 \label{fig:fi2}
\end{figure}

We found C-attractor in this interval of $\alpha$ numerically 
(figures \ref{fig:1} and \ref{fig:fi2}). 
The problem is not trivial because C-attractor 
lays in the unstable manifold (two stable and one unstable directions). 
We use straddle-orbit 
method due to Battelino et al. \cite{battelino} and
H. Nusse and J. Yorke \cite{nusse}.

This procedure allows us to pursue the unstable orbit laying in the
basin boundary in principle forever. It can be viewed as a series
of bisections and goes as follows. At initial time we choose two
points $P_{A}(x=0)$ and $P_{B}(x=0)$ which lead to different attractors A
and B and perform bisection until the distance between the iterates
$P_{A}(0)$ and $P_{B}(0)$ is less a prescribed $\delta$. Next we
integrate the equations numerically starting from the current 
$P_{A}(0)$ and $P_{B}(0)$ until the distance between the trajectories exceeds
$\delta$. When this happens at some time $x$ we stop the integration,
assign the points $P_{A}(x)$ and $P_{B}(x)$ as current representatives and
repeat the bisection. Iterating this procedure one can progressively
construct a trajectory staying within a distance $\delta$ from the
codimension one stable manifold.

\begin{figure}
\begin{center}
\includegraphics[angle=0, width=12cm]{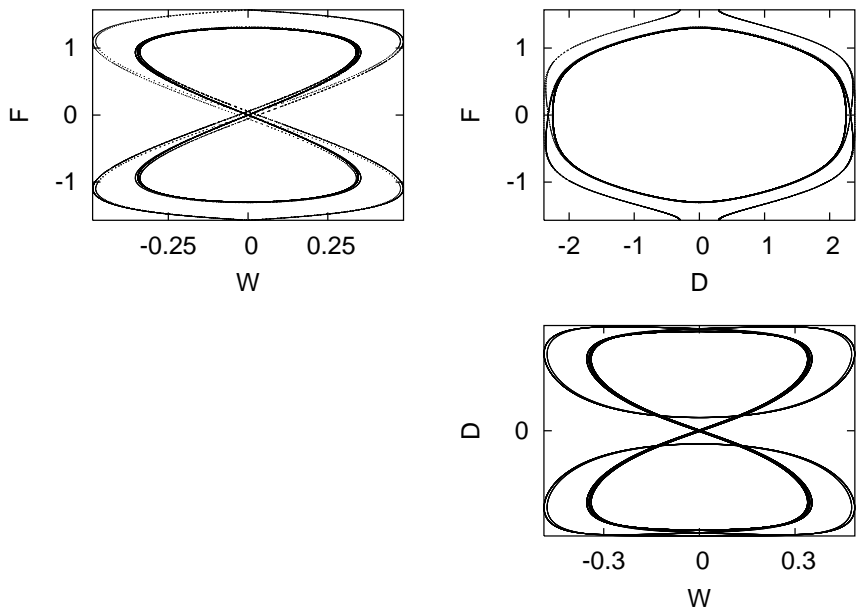}
\end{center}
\caption{The projection of C-attractor, \hbox{$\alpha=0.43$}
(for $F$ we identify the intervals 
$[-\frac{\pi}{2}+\pi k,\frac{\pi}{2}+\pi k]$ where
$k\in N$). Some similarities to the Lorenz attractor are visible.}
\label{fig:3}
\end{figure}

\begin{figure}
\begin{center}
\includegraphics[angle=0, width=12cm]{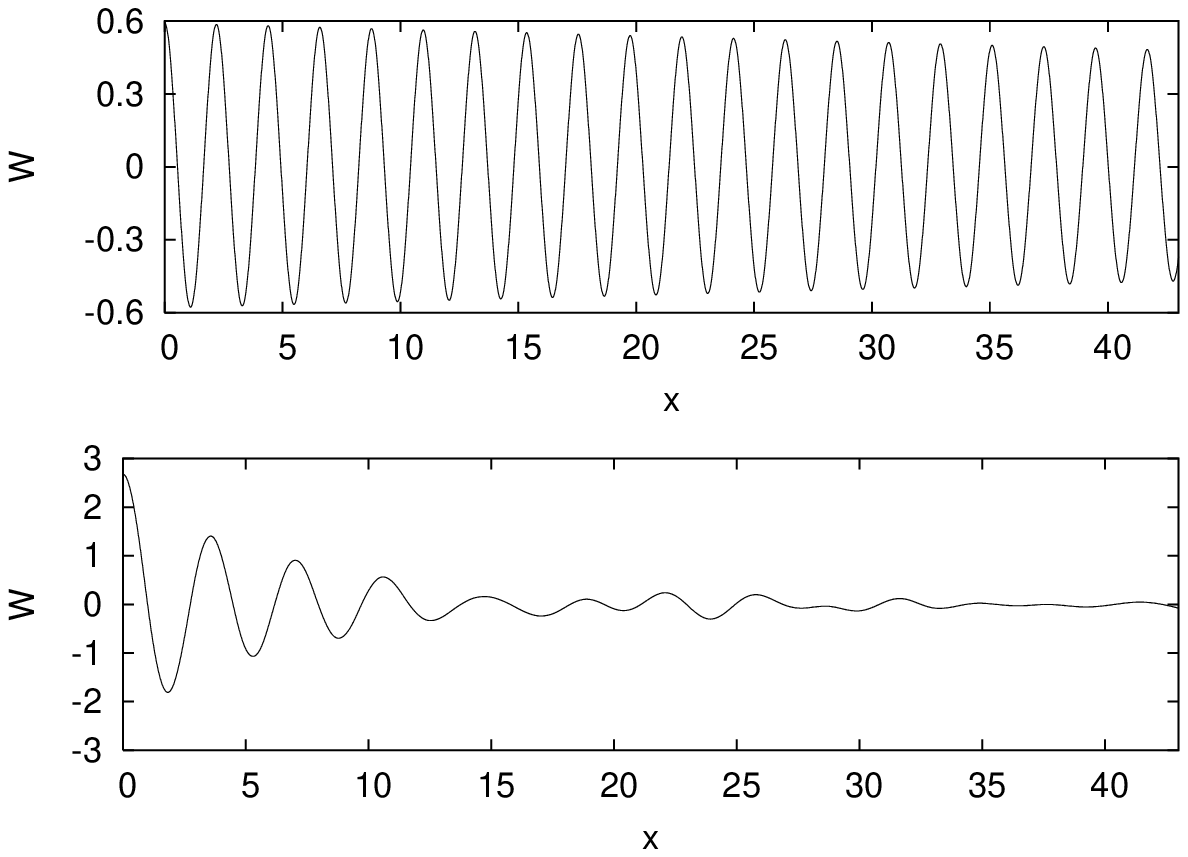}
\end{center}
\caption{The correlation function; the comparison of the periodic solution
\hbox{$\alpha=0.42$}
with chaotic \hbox{$\alpha=0.43$} (only short time correlation exist).}
\label{fig:cor}
\end{figure}

Using straddle-orbit method we found two loops and a sequence of the windings 
(figure \ref{fig:3}). It is in a way similar to the Lorenz attractor but
the nature of C-attractor is different. 
The standard methods like
Lyapunov exponents, test of the subsequent maxima, Poincar\'e sections 
of the attractor do not give good insight into the problem although
behavior of maximal Lyapunov exponent
indicate the chaotic region (see figure 7 in
\cite{periodic_wm}). The best evidence of chaotic properties provides 
the autocorrelation function which exhibits
a typical beha\-vior (figure \ref{fig:cor}). It shows that we are able to
predict short range behavior of the system (inside single loops) with 
good accuracy but long, deterministic sequence of windings 
is uncorrelated. We can show this numerically up to almost 
any finite period with
help of the symbolic dynamics and information theory. 
Symbolization\footnote{It will be described elsewhere.} 
allows us to check periodicity
of our solutions, calculate Shannon entropy or EMC (Effective Measure of
Complexity given by Grassberger \cite{grassberger}) and show sensitivity to
initial conditions.

\section{The horseshoe-like dynamics}

\begin{figure}
\begin{center}
\includegraphics[angle=0, width=12cm]{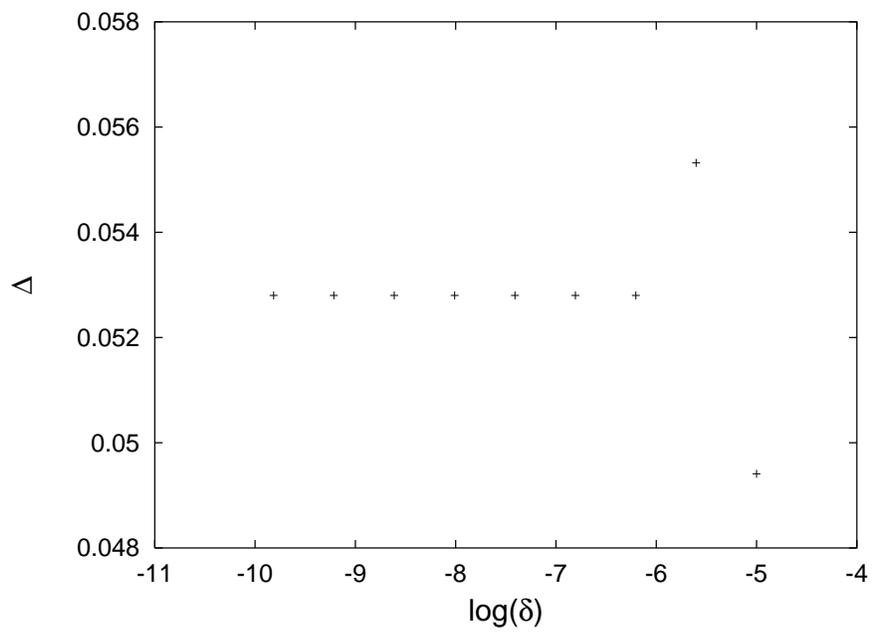}
\end{center}
\caption{The size of chaotic window as a function of the numerical
precision. Since the result does not tend to zero with increasing machine
precision we believe that the observed chaotic behavior is not 
a numerical artifact.}
\label{fig:bif1b}
\end{figure}

The results presented in previous subsections are based on numerical
calculations. It is well known that chaos and fractal structures could appear
in dynami\-cal systems due to the discretization errors \cite{sensitivity}. 
In order to make confident that our result is not a numerical artifact
we checked (in the bifurcation diagram) the dependence of
the size of chaotic windows on machine precision. We have performed this 
calculation with positive result: 
the size of the chaotic windows does not 
tend to zero with increasing machine precision (figure \ref{fig:bif1b}). 
Unfortunately there is not enough computational efficiency 
to use more detailed lattice and estimate the limit.

In summary, we have strong numerical evidence for chaos
but we need a qualitative argument. Here enters the horseshoe.

Now, we will show that there is two dimensional section of the basin boundary
which has a fractal structure. 

\begin{figure}[!h]
\begin{center}
\input{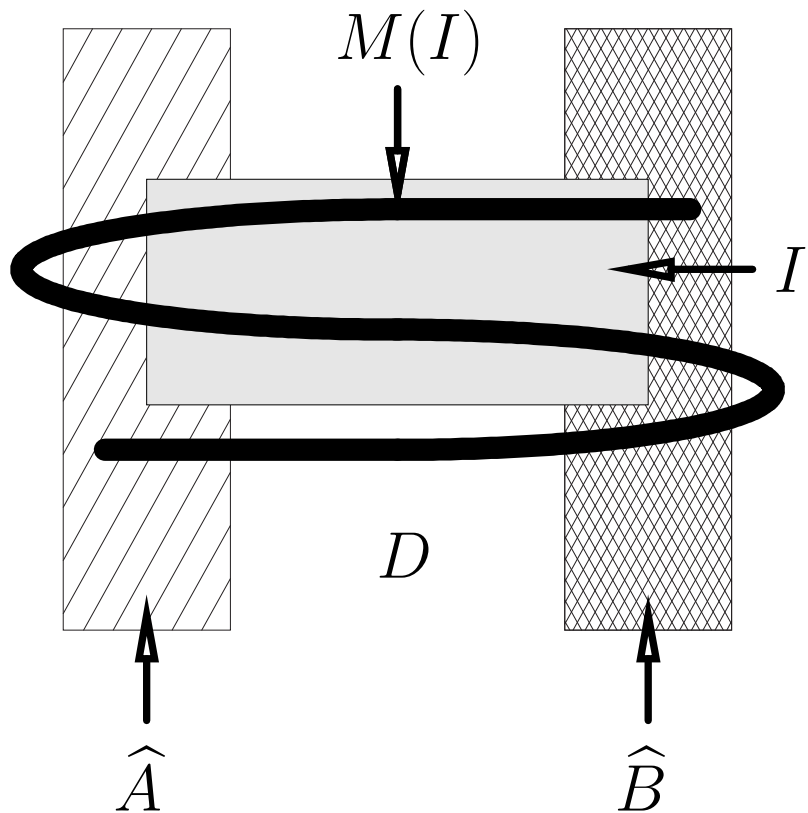}
\caption{Sets: $\widehat{A}$, $\widehat{B}$, test set $I$ and 
$M(I)$ (S-shape). We have:  $\widehat{A}\subset A$ 
and $\widehat{B} \subset B$ where $A$, $B$ are two basin sets.
The horseshoe in the crossing $M(I)$ and $I$ signals fractal boundary
between $A$ and $B$.}
\label{fig:figb2}
\end{center}
\end{figure}

\begin{figure}[!h]
\begin{center}
\input{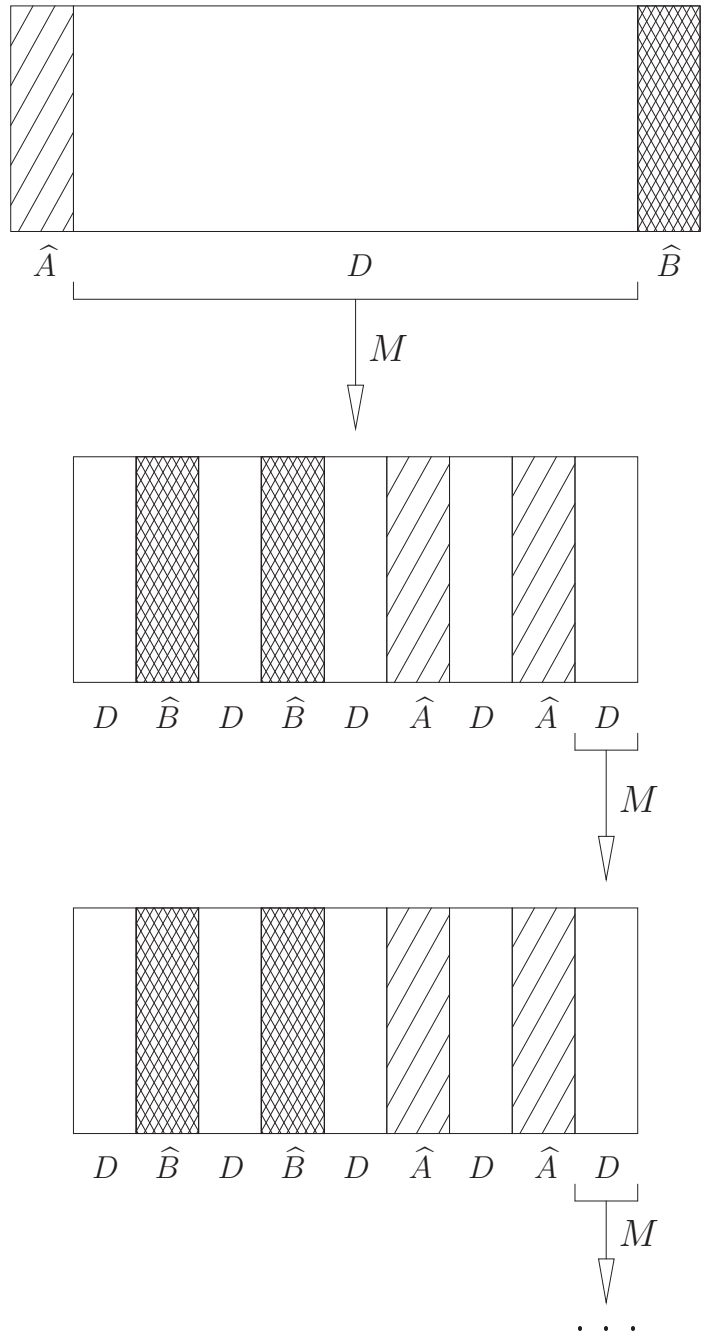}
\caption{The structure of the set $I$ determined after 0, 1, 2 iterations of
the map $M$. The repetitions of this procedure converge to a part of 
the boundary (between basin sets $A$ and $B$) and will lead
to a fat Cantor set in the limit.}
\label{fig:figb3}
\end{center}
\end{figure}

Let us consider a two dimensional one to one and continuous map $M$. 
We choose two
sets $\widehat{A}$ and $\widehat{B}$ such that $\widehat{A}\subset A$ 
and $\widehat{B} \subset B$ where $A$, $B$ are two basin sets. 
Of course we have 
$\widehat{A}\cap \widehat{B}=\oslash$. We choose a test set $I$
(gray rectangle in figure \ref{fig:figb2}) and iterate $M(I)$. 
If $M(I)$ and $I$ cross
in the way presented in figure \ref{fig:figb2} we could say something 
more about structure of set $I$ (see figure \ref{fig:figb3}). 
There are subsets in $I$ which go
to $A$ or $B$. By $D$ we denote this part of domain for which we are not 
sure where it goes to. The repetition of this procedure will reveal more and
more complicated structure of the test set. In the limit we will get
a structure similar to a fat Cantor set.

\begin{figure}[!h]
\begin{center}
\includegraphics[angle=0, width=12cm]{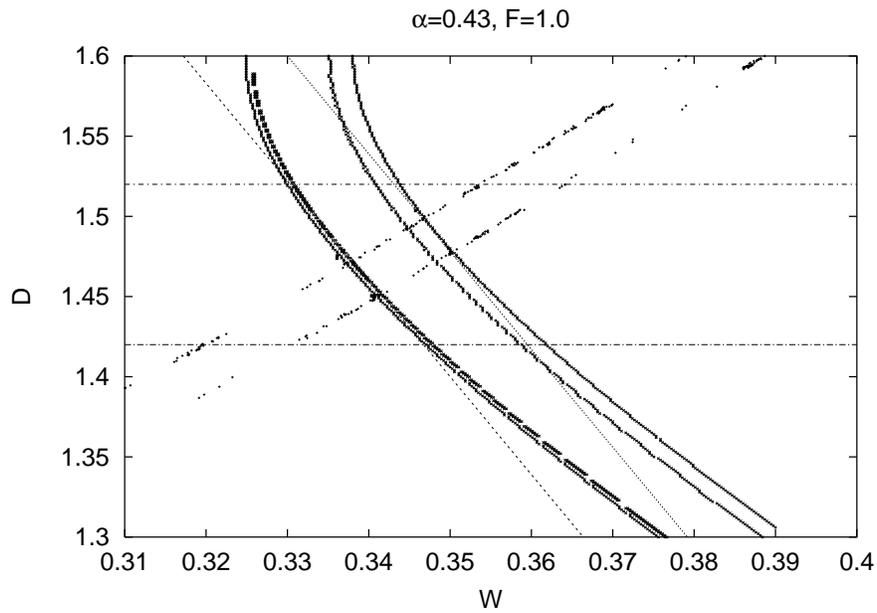}
\end{center}
\caption{The intersection of the test set $I$ (bounded by straight lines)
and its first image by Poincar\'e map. We have plotted in the
background the basin boundary (visible as four, almost parallel curves).
These curves correspond to the stable manifolds of the periodic points.
The first iteration of the test set $I$ corresponds to the unstable manifolds.
We observe horseshoe as in figure \ref{fig:figb2}. This is a ``first 
approximation'' of the heteroclinic intersection (figure \ref{fig:stabun}).}
\label{fig:fi8}
\end{figure}

With the help of the Poincar\'e map we could reduce three dimensional flow 
to two dimensional map (Poincar\'e map is one to one and continuous as 
map $M$).
Test set and its intersection satisfying described 
conditions are presented in figure \ref{fig:fi8}.

It follows from the foregoing that 
there is a surface in the phase space whose intersection with the 
basin boundary has a fractal structure. This, together with numerical data
implies 
existence of the chaotic solution. We believe the mathematically rigorous 
proof can be done by applying the method presented in 
\cite{mischaikow01chaos}.

\section{Route to chaos}

There is also a parallel description of the phenomenon presented in the 
previous
section which helps to follow  ``the route to chaos'' in
Poincar\'e picture. 

\begin{figure}[!h]
\begin{center}
\includegraphics[angle=0, width=12cm]{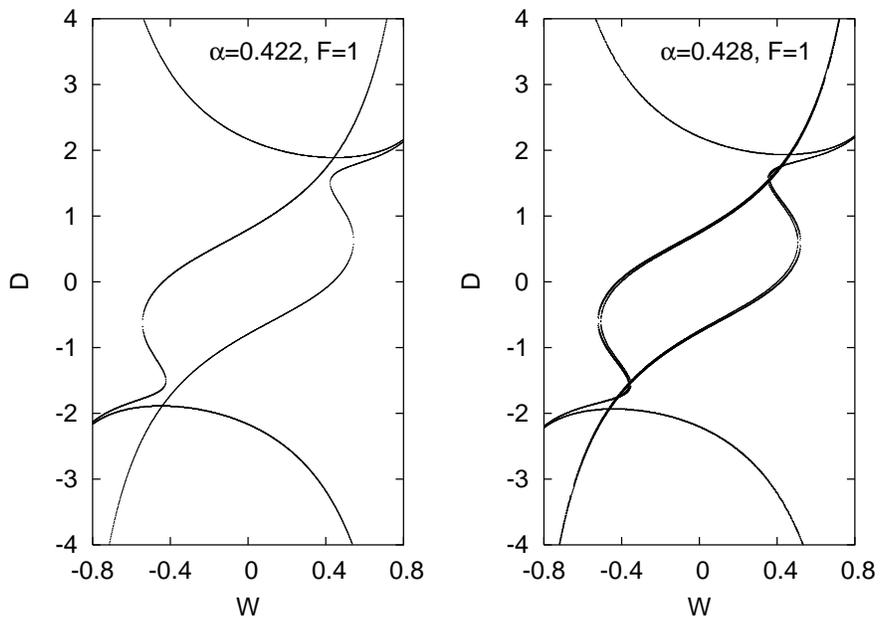}
\end{center}
\caption{Two saddles and the creation of the heteroclinic 
intersections of stable and unstable manifolds.}
\label{fig:stabun}
\end{figure}

The system (\ref{eq:1a}--\ref{eq:3a}) has reflection symmetry
$F\rightarrow -F$ so for each initial value of $c^{*}$ also $-c^{*}$ gives
critical solution. We take a slice of the whole phase space and
we denote by $N$ the map defined by subsequent intersections of the flow
with the chosen slice. This corresponds to four period two points 
for small positive $\alpha$ (intersection of two loops with the surface). 
Map $N^{2}$ reduces four periodic points to two saddles and gives us
proper Poincar\'e section.
These saddles lay in the basin boundary. The
intersection of the basin boundary with Poincar\'e
surface gives stable manifolds of the saddles for $N^{2}$. 
If we choose Poincar\'e
section with $F=const$, we can use
existing symmetries to determine unstable manifolds. The derivatives $W'$,
$D'$ are invariants of the reflection $W\rightarrow -W$, $D\rightarrow -D$.
It follows from this that stable manifolds under 
this transformation correspond to
unstable manifolds (see figure \ref{fig:stabun}). The crossings determine
positions of the saddles. The shapes of the manifolds and 
positions of the saddles
depends on the parameter $\alpha$. 

For small $\alpha$,
$b-orbits$ have been found analytically (see \cite{periodic_wm}).)
We know from this three dimensional analysis that
for $\alpha\rightarrow 0$ periodic solutions
run towards infinity (as we can see in figure \ref{fig:bif_b}) and their
period shrinks to zero. That means in terms of the Poincar\'e picture that
the distance between saddles is unbounded for $\alpha\rightarrow 0$.

If we go in the opposite direction (growing $\alpha$) 
the distance between the unstable
manifold of the first saddle and the stable manifold of the second saddle
decreases. At the bifurcation point they collide ($\alpha\simeq 0.426$) 
and the heteroclinic
intersection arises (see figure \ref{fig:stabun}). This implies an
infinite number of intersections and a chaotic nature of C-attractor.
For larger values of $\alpha$ ($\alpha\simeq 0.482$) the heteroclinic
intersection disappears and C-attractor becomes periodic again.

\begin{figure}
\begin{center}
\includegraphics[angle=0, width=12cm]{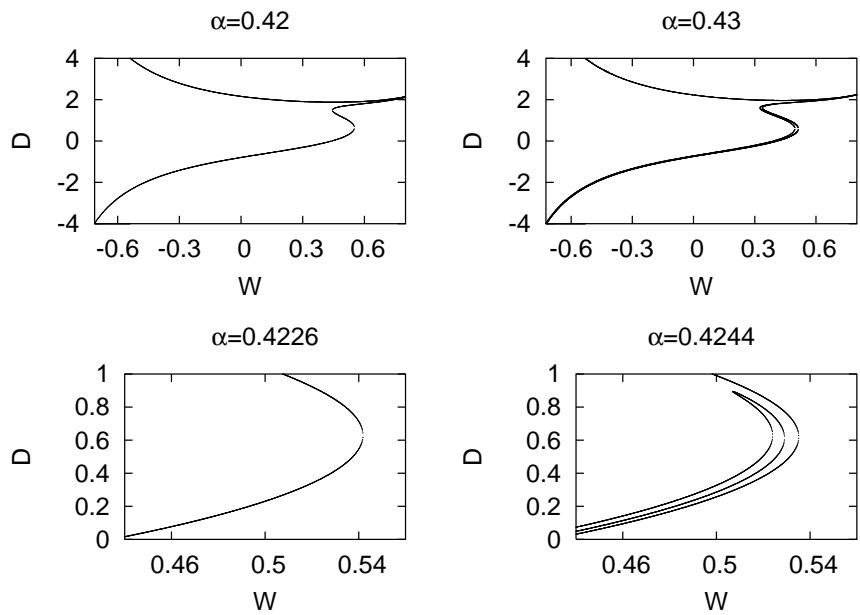}
\end{center}
\caption{The bifurcation of the basin boundary, F=1.0.
The strong deformation is observed ($\alpha=0.4244$)
before a chaotic behavior appears ($\alpha\simeq 0.426$).}
\label{fig:4b}
\end{figure}

The bifurcation of the basin boundary seems to be different 
from the one ''at point'' described for the Henon map.
We have observed that manifolds are deformed just before bifurcation
point ($\alpha=0.4244$, figure \ref{fig:4b}). Further investigation
is needed to clarify this fact.

\section{Final remarks}

The main result of this paper is the qualitative and 
quantitative argument for
the existence of a chaotic solutions of the equations 
(\ref{eq:1}-\ref{eq:3}).
We have skipped the physical background but we point out that
these equations are reduced Einstein's equations. In addition,
chaotic solutions separate two generic types so in this sense they
are critical. As far as we know this is the first example
of fractal critical behavior in the context of Einstein's equations
\cite{gundlach}.

\section{Acknowledgments}

I would like to express my gratitude to Piotr Bizo\'n for
help.
 
The research of the author was supported in part by the KBN grant {no. 
2P03B00623}
and in part by the FWF grant \hbox{no. P15738}.

\newpage
\bibliography{art2d}
\bibliographystyle{unsrt}

\newpage

\newpage

\newpage

\newpage

\newpage

\newpage

\newpage

\newpage

\newpage

\newpage

\newpage

\newpage

\newpage

\newpage

\newpage

\newpage

\newpage

\end{document}